\documentstyle[aps]{revtex}
\def\beq{\begin{eqnarray}}
\def\eeq{\end{eqnarray}}
\begin{document}
\title{EPR Correlations as an Angular Hanbury-Brown---Twiss  Effect}
\draft
\author{A. F. Kracklauer\cite{afk}}
\address{Institut f\"ur Mathematik und Physik, Bauhaus Universit\"at, 
Weimar, Deutschland}
\date{\today}
\twocolumn[           
\maketitle
\begin{minipage}{\textwidth}   
\begin{quotation}           
\begin{abstract}
It is shown that EPR correlations are the angular analogue to the 
Hanbury-Brown---Twiss effect.  As insight provided by this model, it 
is seen that, the analysis of the EPR experiment requires conditional 
probabilities which do not admit the derivation of Bell inequalities. 
\end{abstract} 
 \end{quotation}      
\end{minipage}]      

Bell's Theorems purport to prove that an objective, local 
hidden-variable extension of Quantum Mechanics (QM) is impossible.  
This result has been called ``beautiful'' and the century's most 
significant discovery.  For some, however, this result is a symptom 
of error or misunderstanding.  

Of course, a theorem does not establish a universal, unrestricted 
truth; it only tests symbolic manipulations, that is mathematics, for 
consistency, {\it given an hypothesis}.  A search for error in Bell's 
analysis, therefore, is nothing but a critical review of its 
hypothesis.  The obdurate realist, who wishes to challenge Bell's 
conclusion, has only two options: QM must be wrong (perhaps incomplete 
or otherwise defective on the margins) or, the hypothesis contains 
error.  

Within QM, all that is needed to obtain the expressions relevant to 
the Einstein, Rosen and Podolsky (EPR) experiment, as modified by 
Bohm (EPRB), at the heart of Bell's analysis, is a superposition 
state; the rest follows from simple geometric transformations.  
Indeed, exactly this feature of QM has been questioned, starting with 
Furry.\cite{Furry}  He suggested that for macroscopic distances, a 
superposition state converts to a mixed state. 

The second option, seeking error in the hypothesis of what should 
constitute an objective local extension of QM, is likewise lean 
on possibilities.  The hypothesis Bell used was scarcely more than the 
assertion that the coincidence intensity for the EPRB experiment is 
to be given by: 
\beq P(a,b)=\int 
I_{\cal A}(a,\lambda) I_{\cal B}(b,\lambda) \rho (\lambda) d\lambda, 
\eeq 
where notation and content are taken from Bell with the modification 
that $I_A$ stands for the count rate, or  photoemission probability, 
at measuring station $A$, etc.\cite{Bell} For ideal photodetectors, 
this count rate is proportional to the impinging field intensity; 
i.e., to the square of the field strength. 

It is the purpose here to analyze just these assumptions, in 
particular the second, and to show that in fact application of the 
principles underlying the Hanbury-Brown---Twiss Effect, permits an 
objective, local interpretation of the EPR correlations. 

The application of Furry's proposal to the EPRB experiment, proceeds as 
follows.  It is assumed that the source emits classical 
electromagnetic radiation polarized in a particular but random 
direction.  It is taken that in each arm of the setup, this radiation 
is to be directed through a polarizer and then detected using a 
photodetector which obeys the square law; i.e., it emits 
photoelectrons in proportion to the square of the intensity of the 
absorbed radiation.  That is, the probability of emission of a 
photoelectron in each arm of an EPR experiment is $({\cal 
E}\cos(\theta))^2$ where $\theta$ is the angle between the 
polarization direction of the signal and the axis of the polarizer 
used in the detector.  A coincidence detection is then taken to be 
proportional to the pro\-duct of detection probabilities in each 
channel,  $\cos^2(\theta)\cos^2(\theta -\phi)$, where $\phi$ is 
the angle between the axes of the measurement polarizers if the 
coordinate system is aligned with one of them. That this 
product gives a coincidence probability is based on the proposition 
that the probability of coincidence of local and therefore 
statistically independent events is the product of the individual 
probabilities.  In the notation of Equation~1, $I_{\cal A} = 
\cos^2(\theta),\,\, I_{\cal B} = \cos^2(\theta - \phi) $and$ 
\rho(\lambda) = d\lambda /2\pi$. 

Finally, the total coincidence rate is obtained by averaging over many 
pairs of signals, each with its own randomly given polarization angle 
$\theta$, that is 
\begin{eqnarray} 
{1\over{2\pi}}\int_{0}^{2\pi}[\cos(\theta)\cos(\theta - \phi)]^2 
d\theta = 1/4 +{1/8}cos(2\phi). 
\end{eqnarray} 
To convert this intensity to a probability it must be recast as  
the ratio of a coincidence rate divided by the total count rate.  For 
ideal detectors, the total number of detections is linearly 
proportional to the sum of the field intensities at both detectors, 
that is: $2(\int_0^{2\pi}I(\theta)d\theta) = 1$.
This model was examined as a semiclassical EPR variant in Ref. [3]. 

This expression seems perfectly rational and, as the resulting 
correlation is $cos(2\phi)/2$, it does not violate a Bell Inequality.  
It would resolve the conundrums evoked by Bell's Theorems were it to 
agree with experiment.  However, this result has a nonzero minimum, 
whereas the QM equivalent, $\cos^2(\phi)/2$, does go to zero and this 
difference has been observed and reported in Ref. [3].  Eq.~(2) does 
not conform to Nature. 

If Furry's {\it Ansatz} is to benefit a realist program, it must 
therefore be modified in some essential.  The search need not be 
carried far.  The radiation in an EPR experiment emanates from a 
single source (which may comprise many microunits, atoms say) and is 
by assumption such that the twin emissions pairwise are not to carry 
off angular momentum.  In the case of radiation, this effectively 
means that if one is polarized so as to pass a polarizer in any 
particular direction, the other must be blocked by a polarizer in 
this direction.  Obviously, such emissions are not {\it statistically 
independent} in each arm.  In turn, without statistical independence, 
the assumption of factorizability of the joint probability as employed 
in writing Eq. (1), is not admissible.  Factorization is overly 
restrictive and not valid.\cite{Mandel}   Quite reasonably so, as the 
question is: given that a particular result is obtained in one arm, --
- absolute simultaneity is impossible --- what are the probabilities 
of outcomes in the second arm?  This calculation demands conditional 
probabilities and they are not factorizable.  

This in no way, however, implies nonlocality; rather, it implies just 
{\it statistical dependence}; i.e., the probabilities of `Bertlmann's 
socks.'  Nonlocality, taken as a violation of Einstein's principle 
that all influences effective at a particular event (point) in 
Minkowski space, must originate at points in the past light cone of 
that event, is not violated.  The correlation resulting from most 
forms (a realist holds: all forms) of statistical dependence is simply 
derived from a common cause.  Of course, as factorizability always 
holds for the coincidence probability of statistically independent 
events, it inevitably implies no violation of locality.  Indeed, such 
events have no cause-effect relationship.  In summary, factorizable 
coincidences map onto but are not one-to-one on the set of all 
coincidence functions for events respecting locality.\cite{comment}

Thus, as an alternate to the Furry inspired model described above, 
consider the following:

\begin{description} 
\item[a.\phantom{12345}] The source is assumed to emit in the 
$\pm{\hat{z}}$ direction circularly polarized signals; clockwise in 
one direction and counterclockwise in the other.  Thus, the signal 
impinging on photodetector A, say, is: 
\begin{eqnarray}
E_A(\theta)=\hat{x}cos(\theta) + e^{i\pi/4}\hat{y}sin(\theta),
\end{eqnarray}
where factors of the form $exp(k\cdot x -\omega t)$ are supressed, 
$\hat{x},\hat{y}$ are orthogonal unit vectors, the factors 
$cos(\theta), sin(\theta)$ project the individual components of the 
circularly polarized signal onto the axis of the polarizer and the 
factor, $ exp(i\pi/4),$ represents the fixed phase difference between 
the orthogonal components which give circular polarization.   
Likewise, the signal impinging on the photodetector B, oriented at 
angle $\phi$ with respect to A, as expressed in $A$'s coordinate 
system, is: 

\begin{eqnarray}
E_B(\theta, \phi)=\hat{y}cos(\theta -\phi) - 
e^{i3\pi/4}\hat{x}sin(\theta -\phi). \end{eqnarray} 

\item[b.\phantom{12345}] Use is made now of a generalized coincidence 
probability inspired by second order coherence theory:

\begin{eqnarray} P(a,b) = {\langle E_A\cdot E_B E_B\cdot E_A \rangle
\over{\langle |E_A|^2 + |E_B|^2\rangle}}, \label{coin}  \end{eqnarray} 
where the angle brackets indicate an ensemble average over all values 
of $\theta$, the angle of attack of each separate signal, or, on an 
ergotic principle, over the random phases of the individual atomic 
sources. The dot product is with respect to the orthogonal set 
$\{\hat{x},\hat{y}\}$.  

The numerator in Equation \ref{coin} is the probability of a 
coincidence count; as usual for ideal detectors, it is the product of 
the intensities of the separate signals, but in the form taught by 
coherence theory.  Traditionally, intensity correlation calculations 
were based on the direct product of intensities, $\langle I_1 I_2 
\rangle$, whereas coherence theory teaches that the correct form for 
this calculation is $\langle E_1\cdot E_2 E_2\cdot E_1 
\rangle$.\cite{Mandel} The effective difference is that the later form 
allows the phase to contribute to calculation.  It is the information 
in the phase that is required to explain the Hanbury-Brown---Twiss 
effect as well as other coherence phenomena. 

Note that all the information used in the calculation of the numerator 
of Equation \ref{coin} is propagated to stations $A$ and $B$ from 
events in the past light cones of these events.  The signals arriving 
at the measurement stations are, in this case, just classical 
electromagnetic signals for which there is no question of a violation 
of locality.  Here it is seen clearly that factorizability is not a 
valid encodification of Einsteinian locality.

The denominator in Equation \ref{coin} is equal to the total intensity 
of both signals in both detectors and is, therefore, proportional to 
the total photoelectron count, again, for ideal detectors.  The ratio 
of the numerator to the denominator then is by definition the 
probability of coincidence counts. \end{description} 

Taking all the above into account, provides the following expression 
for the coincidence count rate:
\begin{eqnarray}
P(\hat{\bf a},\hat{\bf b})=& \nonumber \\ 
& 
{{\int_0^{2\pi}(cos(\theta)sin(\theta -\phi) - 
sin(\theta)cos(\theta -\phi))^2 d\theta} 
\over{2 \int_0^{2\pi}(cos^2 (\theta) + sin^2 (\theta)) d\theta}}.
\end{eqnarray}
Evaluated, this integral equals the QM result, Eq. (3): 
\begin{eqnarray} 
P(\hat{\bf a},\hat{\bf b}) = {1\over{2}}\sin^2(\phi). 
\end{eqnarray} 

This model, comprising non quantum components, is fully local in 
the Einsteinian sense; and, as it agrees with QM, it is in accord with 
those laboratory observations verifying QM.\cite{Clauser} In essence 
it is, given the vector character of electromagnetic radiation, just 
the angular analogue of the Hanbury-Brown---Twiss Effect.  It stands 
as a counterexample to Bell's conclusion. 

Like QM, however, it violates Bell Inequalities.  Such inequalities, 
however, are derived under the assumption that the relevant coincidence 
probabilities factor into two terms, which a second order coherence 
function does not in general allow.\cite{Mandel}  That is: Bell 
inequalities are not valid for all forms of fully local coincidences.

This can also be shown as follows.  As a matter of fact, the most 
general form for the coincidence count between stations $A$ and $B$ in 
the EPRB experiment is a function of three sets of variables: $P({\bf 
a, b}, \lambda)$, where ${\bf  a, b}$ are those variables that specify 
conditions at the measuring stations $A$ and $B$, and $\lambda$ 
specifies all common causes pertaining to the generation of the two 
signals.  The $\lambda$, not being explicit in QM, have been denoted 
``hidden variables.'' The coincidence count considered in QM is the 
marginal probability derived from the full coincidence probability by 
integration over $\lambda$: 
\beq P({\bf a, b}) = \int P({\bf a, b}, \lambda) d\lambda. \eeq 
Now, the identity from probability theory: \beq P({\bf a, b}, \lambda) 
=P(\lambda)P({\bf a}| \lambda)P({\bf b| a,}\,\lambda), \label{pravda} 
\eeq where $P(x| y)$ is the conditional probability of $x$ contingent 
on $y$, exposes the intrinsic structure of such a 
coincidence.\cite{Feller} 

This form reduces to that of the integrand of Equation~1 when $P({\bf 
b|a},\lambda) =  P({\bf b}| \lambda)$; i.e., when the events at $A$ and 
$B$ are {\it statistically independent}; in other words, when there is 
no relationship between them.  This is fundamentally contrary to the 
structure of the EPRB experiment in which it is taken that the 
emissions are correlated.  It is easy to verify that no derivation of 
a Bell inequality goes through using Equation \ref{pravda}.  Thus, 
such inequalities do not pertain to correlated events. 

Equation \ref{pravda} does not imply that information is telegraphed 
from station $A$ to station $B$.  It means only that the counts 
registered at both stations will exhibit correlations that will become 
evident when the data is brought together at a later time for 
comparison.  Such a comparison can be made, naturally, only at a 
point in Minkowski space for which the the past light cone includes 
the measuring stations $A$ and $B$.  Likewise, the correlations did 
not arise with the help of superluminal, or any other, communication.  
The structure yielding the correlations when the measuring stations 
are specified by ${\bf a, b}$, is built into these signals at their 
source which is in the past light cones of both stations.  $P({\bf b}| 
{\bf a}, \lambda)$ being contingent on a is a realization not of 
communication between stations $A$ and $B$, but of correlations 
invested in the signals at the common source.  For the EPRB 
experiment, clearly, there can be no coincident count when the 
polarizers are parallel regardless of the orientation of the signals 
(so long as they are orthogonal, as assumed in the first place).  
Thus, the dependence of the conditional probability is the consequence 
of the necessity of the detectors to be set so as to admit detection 
of the correlated characteristics of the signals, here orthogonal 
polarization. 

Of additional interest is the fact that the new model moves the 
nonfactorizable structure from a superposition wave function to the 
form of the coincidence probability.  This affords considerable 
simplification of discussions on the interpretation of QM.  
Superposition wave functions have been the source of much confusion, 
requiring as they do,  ``collapse'' for ontological meaning. 

In conclusion, using the correct classical-physics method to calculate 
a coincidence count in the EPRB experiment, yields the QM result and 
exposes an inappropriate assumption in the derivation of Bell 
inequalities.  No error has been found in QM, rather, just the argument 
against an objective local extension of QM has been put aside.  This 
is at no cost to any established theoretical or empirical result from 
QM.  The fact that experiments to test Bell inequalities have 
virtually beyond all argument supported QM, do not by themselves imply 
that QM is nonlocal.  They prove no more than that inequalities that 
should obtain for objective local extensions of QM, but were derived 
under a false premise in any case, are not valid.  Indeed, we see, 
they can not be. 

\end{document}